\documentclass[aps,prd,onecolumn,showpacs,showkeys,amsmath,amssymb]{revtex4}
\usepackage{amsfonts}
\usepackage{graphicx}
\usepackage{dcolumn}
\usepackage{bm}
\usepackage[dvips]{color}
\begin{document}
%
%
\title{Light Scalar Meson $\sigma(600)$ in QCD Sum Rule with Continuum }
\author{Hua-Xing Chen$^{1,2}$}
\email{chx@water.pku.edu.cn}
\author{Atsushi Hosaka$^{2}$}
\email{hosaka@rcnp.osaka-u.ac.jp}
\author{Hiroshi Toki$^{2}$}
\email{toki@rcnp.osaka-u.ac.jp}
\author{Shi-Lin Zhu$^{1}$}
\email{zhusl@pku.edu.cn} \affiliation{$^1$Department of Physics and
State Key Laboratory of Nuclear Physics and Technology,
Peking University, Beijing 100871, China \\
$^2$Research Center for Nuclear Physics, Osaka University, Ibaraki,
Osaka 567--0047, Japan}
\begin{abstract}
The light scalar meson $\sigma(600)$ is known to appear at low
excitation energy with very large width on top of continuum states.
We investigate it in the QCD sum rule as an example of resonance
structures appearing above the corresponding thresholds. We use all
the possible local tetraquark currents by taking linear combinations
of five independent local ones. We ought to consider the $\pi$-$\pi$
continuum contribution in the phenomenological side of the QCD sum
rule in order to obtain a good sum rule signal. We study the
stability of the extracted mass against the Borel mass and the
threshold value and find the $\sigma(600)$ mass at 530 MeV $\pm$ 40
MeV. In addition we find the extracted mass has an increasing
tendency with the Borel mass, which is interpreted as caused by the
width of the resonance.
\end{abstract}
\maketitle
\pagenumbering{arabic}
%
%

\section{Introduction}\label{sec:intro}

The light scalar mesons, $\sigma(600)$, $\kappa(800)$, $f_0(980)$
and $a_0(980)$, have been intensively discussed for many
years~\cite{Amsler:2004ps,Bugg:2004xu,Klempt:2007cp}. However, their
nature is still not fully
understood~\cite{Caprini:2005zr,Hatsuda:1994pi,Oller:1997ti,Sugiyama:2007sg,Prelovsek:2010gm}.
They have the same quantum numbers $J^{PC}=0^{++}$ as the vacuum,
and hence the structure of these states is a very important subject
in order to understand non-perturbative properties of the QCD vacuum
such as spontaneous chiral symmetry breaking. They compose of the
flavor $SU(3)$ nonet with the mass below 1 GeV, and have a mass
ordering which is difficult to be explained by using a $q \bar q$
configuration in the conventional quark
model~\cite{Amsler:2008zzb,Aitala:2000xu,Ablikim:2004qna,Aston:1987ir,Akhmetshin:1999di}.
Therefore, several different pictures have been proposed, such as
tetraquark states and meson-meson bound states, etc. Here we note
that hadrons with complex structures such as tetraquarks may exist
in the continuum above the threshold energy of two hadrons with
simple quark structure.

The tetraquark structure of the scalar mesons was proposed long time
ago by Jaffe with an assumption of strong diquark
correlations~\cite{Jaffe:1976ig,Jaffe:1976ih}.  It can naturally
explain their mass ordering and decay
properties~\cite{Alford:2000mm,Maiani:2004uc,Weinstein:1990gu}. Yet
the basic assumption of diquark correlation is not fully
established. In this letter, we study $\sigma(600)$ as a tetraquark
state in the QCD sum rule approach as an example of resonances in
the continuum states above the $\pi$-$\pi$ threshold. In the QCD sum
rule, we calculate matrix elements from the QCD (OPE) and relate
them to observables by using dispersion relations.  Under suitable
assumptions, the QCD sum rule has proven to be a very powerful and
successful non-perturbative method in the past
decades~\cite{Shifman:1978bx,Reinders:1984sr}. Recently, this method
has been applied to the study of tetraquarks by many
authors~\cite{Bracco:2005kt,Narison:2005wc,Lee:2006vk,Chen:2007xr}.

In our previous paper~\cite{Chen:2007xr}, we have found that the QCD
sum rule analysis with tetraquark currents implies the masses of
scalar mesons in the region of 600 -- 1000 MeV with the ordering
$m_\sigma < m_\kappa < m_{f_0, a_0}$, while the conventional $\bar q
q$ current is considerably heavier (larger than 1 GeV). To get this
result, first we find there are five independent local tetraquark
currents, and then we use one of these currents or linear
combinations of two currents to perform the QCD sum rule analysis.
But these interpolating currents do not describe the full space of
tetraquark currents. In order to complete our previous study, we use
more general currents by taking linear combinations of all these
currents. It describes the full space of local tetraquark currents
which can couple to $\sigma(600)$. Since $\sigma(600)$ meson is
closely related to the $\pi$-$\pi$ continuum and it has a wide decay
width, we also consider the contribution of the $\pi$-$\pi$
continuum as well as the effect of the finite decay width.

This paper is organized as follows. In Sec.~\ref{sec:qsr}, we
establish five independent local tetraquark currents, and perform a
QCD sum rule analysis by using linear combinations of five single
currents. In Sec.~\ref{sec:numerical}, we perform a numerical
analysis, and we also study the contribution of $\pi$-$\pi$
continuum. In Sec.~\ref{sec:width}, we consider the effect of the
finite decay width. Sec.~\ref{sec:summary} is devoted to summary.

\section{QCD Sum Rule}\label{sec:qsr}

The local tetraquark currents for $\sigma(600)$ have been worked out
in Ref~\cite{Chen:2007xr}. There are two types of currents:
diquark-antidiquark currents $(qq)(\bar q \bar q)$ and meson-meson
currents $(\bar q q)(\bar q q)$. These two constructions can be
proved to be equivalent, and they can both describe the full space
of local tetraquark currents~\cite{Chen:2007xr}. Therefore we shall
just use the first ones. Since we use their linear combinations to
perform the QCD sum rule analysis, we can not distinguish whether it
is a diquark-antidiquark state or a meson-meson bound state.
However, we find that tetraquark currents with a single term do not
lead to a reliable QCD sum rule result which means that
$\sigma(600)$ probably has a complicated structure. The five
independent local currents are given by:
\begin{eqnarray}
\nonumber\label{eq:sigma_current} S^\sigma_3 &=& (u_a^T C \gamma_5
d_b)(\bar{u}_a \gamma_5 C \bar{d}_b^T - \bar{u}_b \gamma_5 C
\bar{d}_a^T)\, ,
\\ \nonumber
V^\sigma_3 &=& (u_a^T C \gamma_{\mu} \gamma_5 d_b)(\bar{u}_a
\gamma^{\mu}\gamma_5 C \bar{d}_b^T - \bar{u}_b \gamma^{\mu}\gamma_5
C \bar{d}_a^T)\, ,
\\
T^\sigma_6 &=& (u_a^T C \sigma_{\mu\nu} d_b)(\bar{u}_a
\sigma^{\mu\nu} C \bar{d}_b^T + \bar{u}_b \sigma^{\mu\nu} C
\bar{d}_a^T)\, ,
\\ \nonumber
A^\sigma_6 &=& (u_a^T C \gamma_{\mu} d_b)(\bar{u}_a \gamma^{\mu} C
\bar{d}_b^T + \bar{u}_b \gamma^{\mu} C \bar{d}_a^T)\, ,
\\ \nonumber
P^\sigma_3 &=& (u_a^T C d_b)(\bar{u}_a C \bar{d}_b^T - \bar{u}_b C
\bar{d}_a^T)\, .
\end{eqnarray}
The summation is taken over repeated indices ($\mu$, $\nu, \cdots$
for Dirac, and $a, b, \cdots$ for color indices). The currents $S$,
$V$, $T$, $A$ and $P$ are constructed by scalar, vector, tensor,
axial-vector, pseudoscalar diquark and antidiquark fields,
respectively. The subscripts $3$ and $6$ show that the diquarks
(antidiquarks) are combined into the color representations,
$\mathbf{\bar 3_c}$ and $\mathbf{6_c}$ ($\mathbf{3_c}$ and
$\mathbf{\bar 6_c}$), respectively.

These five diquark-antidiquark currents $(qq)(\bar q \bar q)$ are
independent. In this work we use general currents by taking linear
combinations of these five currents:
\begin{eqnarray}
\label{eq:eta} \eta &=& t_1 e^{i \theta_1} S^\sigma_3 + t_2 e^{i
\theta_2} V^\sigma_3 + t_3 e^{i \theta_3} T^\sigma_6 + t_4 e^{i
\theta_4} A^\sigma_6 + t_5 e^{i \theta_5} P^\sigma_3\, ,
\end{eqnarray}
where $t_i$ and $\theta_i$ are ten mixing parameters, whose linear
combination describes the full space of local currents which can
couple to $\sigma(600)$.  We can not determine them in advance and
therefore we choose them randomly for the study of the QCD sum rule.

By using the current in Eq.~(\ref{eq:eta}), we calculate the OPE up
to dimension eight. To simplify our calculation, we neglect several
condensates, such as $\langle g^3 G^3\rangle$, etc., and we do not
consider the $\alpha_s$ correction, such as $g^2 \langle \bar q q
\rangle^2$, etc. The obtained OPE are shown in the following. We
find that most of the crossing terms are not important such as
$\rho_{13}$, and even more some of them disappear: $\rho_{15}=0$,
etc. For the most cases, we find that the OPE terms of Dim=6 and
Dim=8 give major contributions in the OPE series in our region of
interest. This is because the condensates $\langle\bar q q\rangle^2$
(D=6) and $\langle\bar q q\rangle\langle g \bar q \sigma G q\rangle$
(D=8) are much larger than others.

\begin{figure}[hbt]
\begin{center}
\scalebox{0.6}{\includegraphics{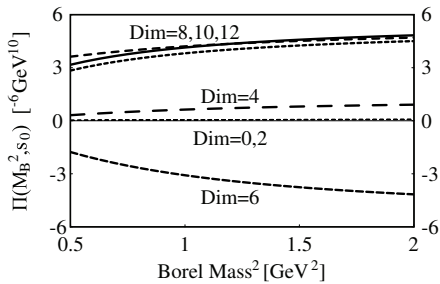}} \caption{The convergence of
the two-point correlation function $\Pi(M_B,s_0)$. The threshold
value is taken to be $s_0 = 1$ GeV$^2$, and we show its behavior up
to certain dimensions, as functions of $M_B^2$. The solid line is
for $\Pi(M_B,s_0)$ up to Dim=8. The short-dashed line around it is
for $\Pi(M_B,s_0)$ up to Dim=10, and the long-dashed line around it
is for $\Pi(M_B,s_0)$ up to Dim=12.} \label{pic:pi}
\end{center}
\end{figure}

Since the OPE series should be convergent to give a reliable QCD sum
rule, we also calculate the OPE of Dim=10 and Dim=12. However, we
find that these terms are not important. Using the parameter set (2)
and the the values of the condensates of the next section as an
example, we show the convergence of the two-point correlation
function $\Pi(M_B,s_0) \equiv \int_0^{s_0} \rho(s) e^{-s/M_B^2} ds$
in Fig.~\ref{pic:pi} as functions of $M_B^2$. The threshold value is
taken to be $s_0 = 1$ GeV$^2$, and we show its behavior up to
certain dimensions. We find that the OPE up to Dim=0 and Dim=2 are
very small; the OPE of Dim=4 gives a minor contribution; the OPE of
Dim=6 and Dim=8 are both important; the OPE of Dim=10 and Dim=12 are
both small, and so we shall neglect them in the following analysis.
%
\begin{eqnarray}
\rho(s) &=& t_1^2 \rho_{11}(s) + t_2^2  \rho_{22}(s) + t_3^2
\rho_{33}(s) + t_4^2 \rho_{44}(s) + t_5^2 \rho_{55}(s)
\\ \nonumber && + 2 t_1 t_2 \cos{(\theta_1 - \theta_2)}
\rho_{12}(s) + 2 t_1 t_3 \cos{(\theta_1 - \theta_3)} \rho_{13}(s) +
2 t_1 t_4 \cos{(\theta_1 - \theta_4)} \rho_{14}(s)
\\ \nonumber && + 2 t_2 t_3 \cos{(\theta_2 - \theta_3)}  \rho_{23}(s)
+ 2 t_2 t_4 \cos{(\theta_2 - \theta_4)} \rho_{24}(s) + 2 t_2 t_5
\cos{(\theta_2 - \theta_5)} \rho_{25}(s)
\\ \nonumber && + 2 t_3 t_4 \cos{(\theta_3 - \theta_4)}
\rho_{34}(s) + 2 t_3 t_5 \cos{(\theta_3 - \theta_5)} \rho_{35}(s) \,
,
\end{eqnarray}
%
where
\begin{eqnarray}
\rho_{11}(s) &=& \frac{s^4} {61440 \pi^6} + (-\frac{{m_u}^2 } {1536
\pi^6} + \frac{{m_u}m_d } {1536 \pi^6} -\frac{{m_d}^2 } {1536
\pi^6}) s^3 + ( \frac{\langle g^2GG \rangle} {6144 \pi^6} -
\frac{{m_u} \langle \bar{q}q \rangle} {192 \pi^4} - \frac{{m_d}
\langle \bar{q}q \rangle} {192 \pi^4}  ) s^2
\nonumber \\
&& + ( - \frac{m_u^2 \langle g^2GG \rangle}{1024\pi^6} + \frac{m_u
m_d \langle g^2GG \rangle}{1024\pi^6} - \frac{m_d^2 \langle g^2GG
\rangle}{1024\pi^6} - \frac{m_u \langle g\bar{q} \sigma Gq
\rangle}{64\pi^4} - \frac{m_d \langle g\bar{q} \sigma Gq
\rangle}{64\pi^4} + \frac{\langle \bar{q}q \rangle ^2}{12\pi^2} ) s
\\
\nonumber && - \frac{7m_u^2 \langle \bar{q}q \rangle ^2}{48\pi^2} +
\frac{m_u m_d \langle \bar{q}q \rangle ^2}{4\pi^2} - \frac{7m_d^2
\langle \bar{q}q \rangle ^2}{48\pi^2} - \frac{m_u\langle g^2 GG
\rangle\langle \bar{q}q \rangle}{768\pi^4} - \frac{m_d\langle g^2 GG
\rangle\langle \bar{q}q \rangle}{768\pi^4} + \frac{\langle \bar{q}q
\rangle \langle g\bar{q}\sigma Gq \rangle}{12\pi^2} \, ,
\end{eqnarray}
\begin{eqnarray}
\rho_{22}(s) &=& \frac{s^4} {15360 \pi^6} +(-\frac{{m_u}^2 } {384
\pi^6}-\frac{{m_u}m_d } {768 \pi^6}-\frac{{m_d}^2 } {384 \pi^6})s^3
+( \frac{\langle g^2GG \rangle} {3072 \pi^6} + \frac{{m_u} \langle
\bar{q}q \rangle} {24 \pi^4} + \frac{{m_d} \langle \bar{q}q \rangle}
{24 \pi^4} ) s^2
\\
&& + ( - \frac{ m_u^2 \langle g^2GG \rangle}{512\pi^6} + \frac{ m_u
m_d \langle g^2GG \rangle}{512\pi^6}- \frac{ m_d^2 \langle g^2GG
\rangle}{512\pi^6} + \frac{m_u \langle g\bar{q} \sigma Gq
\rangle}{32\pi^4} + \frac{m_d \langle g\bar{q} \sigma Gq
\rangle}{32\pi^4} -\frac{\langle \bar{q}q \rangle ^2}{6\pi^2}) s
\\ \nonumber &&
+\frac{11m_u^2 \langle \bar{q}q \rangle ^2}{12\pi^2}+\frac{2m_u m_d
\langle \bar{q}q \rangle ^2}{\pi^2}+\frac{11m_d^2 \langle \bar{q}q
\rangle ^2}{12\pi^2} - \frac{m_u \langle g^2GG \rangle \langle
\bar{q}q \rangle}{384\pi^4} - \frac{m_d \langle g^2GG \rangle
\langle \bar{q}q \rangle}{384\pi^4} -\frac{\langle \bar{q}q \rangle
\langle g\bar{q}\sigma Gq \rangle}{6\pi^2} \, ,
\end{eqnarray}
\begin{eqnarray}
\rho_{33}(s) &=& \frac{s^4} {1280 \pi^6} + (-\frac{{m_u}^2 } {32
\pi^6}-\frac{{m_d}^2 } {32 \pi^6})s^3 +( \frac{11 \langle g^2GG
\rangle} {768 \pi^6} +\frac{{m_u} \langle \bar{q}q \rangle} {4
\pi^4} +\frac{{m_d} \langle \bar{q}q \rangle} {4 \pi^4}  ) s^2 \\
\nonumber && + (- \frac{11 m_u^2 \langle g^2GG \rangle}{128\pi^6} -
\frac{11 m_d^2 \langle g^2GG \rangle}{128\pi^6} ) s \\ \nonumber
&&+\frac{5m_u^2 \langle \bar{q}q \rangle ^2}{\pi^2}+\frac{20m_u m_d
\langle \bar{q}q \rangle ^2}{\pi^2}+\frac{5m_d^2 \langle \bar{q}q
\rangle ^2}{\pi^2} + \frac{11 m_u\langle g^2 GG \rangle\langle
\bar{q}q \rangle}{96\pi^4} + \frac{11 m_d\langle g^2 GG
\rangle\langle \bar{q}q \rangle}{96\pi^4} \, ,
\end{eqnarray}
\begin{eqnarray}
\rho_{44}(s) &=& \frac{s^4} {7680 \pi^6} + (-\frac{{m_u}^2} {192
\pi^6} +\frac{{m_u}m_d} {384 \pi^6} -\frac{{m_d}^2} {192 \pi^6}) s^3
+  \frac{5 \langle g^2GG \rangle} {3072 \pi^6}   s^2
\\
&& + ( - \frac{5 m_u^2 \langle g^2GG \rangle}{512\pi^6} + \frac{5
m_u m_d \langle g^2GG \rangle}{512\pi^6} - \frac{5 m_d^2 \langle
g^2GG \rangle}{512\pi^6} - \frac{m_u \langle g\bar{q} \sigma Gq
\rangle}{16\pi^4} - \frac{m_d \langle g\bar{q} \sigma Gq
\rangle}{16\pi^4} + \frac{\langle \bar{q}q \rangle ^2}{3\pi^2} ) s
\\
\nonumber && - \frac{m_u^2 \langle \bar{q}q \rangle^2} {6\pi^2} +
\frac{8m_u m_d \langle \bar{q}q \rangle^2} {3\pi^2} - \frac{m_d^2
\langle \bar{q}q \rangle^2} {6\pi^2} + \frac{m_u\langle g^2 GG
\rangle\langle \bar{q}q \rangle}{128\pi^4} + \frac{m_d\langle g^2 GG
\rangle\langle \bar{q}q \rangle}{128\pi^4} + \frac{\langle \bar{q}q
\rangle \langle g\bar{q}\sigma Gq \rangle}{3\pi^2} \, ,
\end{eqnarray}
\begin{eqnarray}
\rho_{55}(s) &=& \frac{s^4} {61440 \pi^6} + (-\frac{{m_u}^2} {1536
\pi^6}-\frac{{m_u}m_d} {1536 \pi^6}-\frac{{m_d}^2} {1536 \pi^6}) s^3
+( \frac{\langle g^2GG \rangle} {6144 \pi^6} + \frac{{m_u} \langle
\bar{q}q \rangle} {64 \pi^4} + \frac{{m_d} \langle \bar{q}q \rangle}
{64 \pi^4} ) s^2
\\ && + (
- \frac{m_u^2 \langle g^2GG \rangle}{1024\pi^6}- \frac{m_u m_d
\langle g^2GG \rangle}{1024\pi^6}- \frac{m_d^2 \langle g^2GG
\rangle}{1024\pi^6} +\frac{m_u \langle g\bar{q} \sigma Gq
\rangle}{64\pi^4} +\frac{m_d \langle g\bar{q} \sigma Gq
\rangle}{64\pi^4} -\frac{\langle \bar{q}q \rangle ^2}{12\pi^2}) s
\\ \nonumber &&
+\frac{17m_u^2 \langle \bar{q}q \rangle ^2}{48\pi^2}+\frac{7m_u m_d
\langle \bar{q}q \rangle ^2}{12\pi^2}+\frac{17m_d^2 \langle \bar{q}q
\rangle ^2}{48\pi^2} + \frac{m_u\langle g^2 GG \rangle\langle
\bar{q}q \rangle}{256\pi^4}+ \frac{m_d\langle g^2 GG \rangle\langle
\bar{q}q \rangle}{256\pi^4}  - \frac{\langle \bar{q}q \rangle\langle
g\bar{q} \sigma Gq \rangle}{12\pi^2} \, ,
\end{eqnarray}
\begin{eqnarray}
\rho_{12}(s) &=& (\frac{{m_u}^2} {3072 \pi^6} + \frac{{m_u}m_d}
{1536 \pi^6} + \frac{{m_d}^2} {3072 \pi^6}) s^3 + ( -\frac{{m_u}
\langle \bar{q}q \rangle} {48 \pi^4} - \frac{{m_d} \langle \bar{q}q
\rangle} {48 \pi^4} ) s^2 \\ \nonumber && + (-\frac{m_u \langle
g\bar{q} \sigma Gq \rangle}{32\pi^4} -\frac{m_d \langle g\bar{q}
\sigma Gq \rangle}{32\pi^4} + \frac{\langle \bar{q}q \rangle
^2}{6\pi^2} ) s - \frac{5m_u^2 \langle \bar{q}q \rangle ^2}{12\pi^2}
- \frac{m_u m_d \langle \bar{q}q \rangle ^2}{2\pi^2} - \frac{5m_d^2
\langle \bar{q}q \rangle ^2}{12\pi^2} + \frac{\langle \bar{q}q
\rangle\langle g\bar{q} \sigma Gq \rangle}{6\pi^2} \, ,
\end{eqnarray}
\begin{eqnarray}
\rho_{13}(s) &=& - \frac{\langle g^2GG \rangle} {1024 \pi^6} s^2  +
( \frac{3m_u^2 \langle g^2GG \rangle}{512\pi^6} + \frac{3m_d^2
\langle g^2GG \rangle}{512\pi^6}) s -\frac{m_u\langle g^2 GG
\rangle\langle \bar{q}q \rangle}{128\pi^4}- \frac{m_d\langle g^2 GG
\rangle\langle \bar{q}q \rangle}{128\pi^4} \, ,
\end{eqnarray}
\begin{eqnarray}
\rho_{14}(s) &=&  ( \frac{3m_u^2 \langle g^2GG \rangle}{4096\pi^6} +
\frac{3m_u m_d \langle g^2GG \rangle}{2048\pi^6} + \frac{3m_d^2
\langle g^2GG \rangle}{4096\pi^6} ) s -\frac{m_u\langle g^2 GG
\rangle\langle \bar{q}q \rangle}{128\pi^4}- \frac{m_d\langle g^2 GG
\rangle\langle \bar{q}q \rangle}{128\pi^4}  \, ,
\end{eqnarray}
\begin{eqnarray}
\rho_{23}(s) &=&  ( -\frac{9m_u^2 \langle g^2GG \rangle}{2048\pi^6}-
\frac{9m_u m_d \langle g^2GG \rangle}{1024\pi^6} - \frac{9m_d^2
\langle g^2GG \rangle}{2048\pi^6} ) s + \frac{3m_u\langle g^2 GG
\rangle\langle \bar{q}q \rangle}{64\pi^4} + \frac{3m_d\langle g^2 GG
\rangle\langle \bar{q}q \rangle}{64\pi^4} \, ,
\end{eqnarray}
\begin{eqnarray}
\rho_{24}(s) &=& \frac{\langle g^2GG \rangle} {1024 \pi^6} s^2 + (
-\frac{3m_u^2 \langle g^2GG \rangle}{512\pi^6} - \frac{3m_d^2
\langle g^2GG \rangle}{512\pi^6}) s + \frac{m_u\langle g^2 GG
\rangle\langle \bar{q}q \rangle}{128\pi^4} + \frac{m_d\langle g^2 GG
\rangle\langle \bar{q}q \rangle}{128\pi^4} \, ,
\end{eqnarray}
\begin{eqnarray}
\rho_{25}(s) &=& ( \frac{m_u^2 \langle g^2GG \rangle}{4096\pi^6} +
\frac{m_u m_d \langle g^2GG \rangle}{2048\pi^6} + \frac{m_d^2
\langle g^2GG \rangle}{4096\pi^6}) s - \frac{m_u\langle g^2 GG
\rangle\langle \bar{q}q \rangle}{384\pi^4} - \frac{m_d\langle g^2 GG
\rangle\langle \bar{q}q \rangle}{384\pi^4} \, ,
\end{eqnarray}
\begin{eqnarray}
\rho_{34}(s) &=& (-\frac{{m_u}^2} {256 \pi^6}-\frac{{m_u} m_d} {128
\pi^6} -\frac{{m_d}^2} {256 \pi^6}) s^3 + ( \frac{{m_u} \langle
\bar{q}q \rangle} {4 \pi^4} + \frac{{m_d} \langle \bar{q}q \rangle}
{4 \pi^4} ) s^2 \\ \nonumber && + ( -\frac{15m_u^2 \langle g^2GG
\rangle}{2048\pi^6}- \frac{15m_u m_d \langle g^2GG
\rangle}{1024\pi^6} - \frac{15m_d^2 \langle g^2GG
\rangle}{2048\pi^6} +\frac{3m_u \langle g\bar{q} \sigma Gq
\rangle}{8\pi^4} +\frac{3m_d \langle g\bar{q} \sigma Gq
\rangle}{8\pi^4} - \frac{2\langle \bar{q}q \rangle ^2}{\pi^2} ) s \\
\nonumber && + \frac{5m_u^2 \langle \bar{q}q \rangle ^2}{\pi^2} +
\frac{6m_u m_d \langle \bar{q}q \rangle ^2}{\pi^2} + \frac{5m_d^2
\langle \bar{q}q \rangle ^2}{\pi^2} + \frac{5m_u\langle g^2 GG
\rangle\langle \bar{q}q \rangle}{64\pi^4}+ \frac{5m_d\langle g^2 GG
\rangle\langle \bar{q}q \rangle}{64\pi^4} - \frac{2\langle \bar{q}q
\rangle\langle g\bar{q} \sigma Gq \rangle}{\pi^2} \, ,
\end{eqnarray}
\begin{eqnarray}
\rho_{35}(s) &=& - \frac{\langle g^2GG \rangle} {1024 \pi^6} s^2 + (
\frac{3m_u^2 \langle g^2GG \rangle}{512\pi^6} + \frac{3m_d^2 \langle
g^2GG \rangle}{512\pi^6}) s - \frac{m_u\langle g^2 GG \rangle\langle
\bar{q}q \rangle}{128\pi^4} - \frac{m_d\langle g^2 GG \rangle\langle
\bar{q}q \rangle}{128\pi^4} \, .
\end{eqnarray}

\section{Numerical Analysis}\label{sec:numerical}

To perform the numerical analysis, we use the values for all the
condensates from
Refs.~\cite{Yang:1993bp,Narison:2002pw,Gimenez:2005nt,Jamin:2002ev,Ioffe:2002be,Ovchinnikov:1988gk}:
%
\begin{eqnarray}
\nonumber &&\langle\bar qq \rangle=-(0.240 \mbox{ GeV})^3\, ,
\\
\nonumber &&\langle\bar ss\rangle=-(0.8\pm 0.1)\times(0.240 \mbox{
GeV})^3\, ,
\\
\nonumber &&\langle g_s^2GG\rangle =(0.48\pm 0.14) \mbox{ GeV}^4\, ,
\\ \nonumber && m_u = 5.3 \mbox{ MeV}\, ,m_d = 9.4 \mbox{ MeV}\, ,
\\
\label{condensates} &&m_s(1\mbox{ GeV})=125 \pm 20 \mbox{ MeV}\, ,
\\
\nonumber && \langle g_s\bar q\sigma G
q\rangle=-M_0^2\times\langle\bar qq\rangle\, ,
\\
\nonumber &&M_0^2=(0.8\pm0.2)\mbox{ GeV}^2\, .
\end{eqnarray}
%
As usual we assume the vacuum saturation for higher dimensional
operators such as $\langle 0 | \bar q q \bar q q |0\rangle \sim
\langle 0 | \bar q q |0\rangle \langle 0|\bar q q |0\rangle$. There
is a minus sign in the definition of the mixed condensate $\langle
g_s\bar q\sigma G q\rangle$, which is different with some other QCD
sum rule calculation. This is just because the definition of
coupling constant $g_s$ is
different~\cite{Yang:1993bp,Hwang:1994vp}.

Altogether we took randomly chosen 50 sets of $t_i$ and $\theta_i$.
Some of these sets of numbers lead to negative spectral densities in
the low energy region of interest, which should be, however,
positive from their definition. This is due to several reasons. One
reason is that the convergence of OPE may not be achieved yet for
those currents for the tetraquark state. Another reason is that some
currents may not couple to the physical states properly. Except
them, there are fifteen sets which lead to positive spectral
densities. We show these fifteen sets of $t_i$ and $\theta_i$ in
Table~\ref{table:parameters}, and label them as (01), (02),
$\cdots$, (15). They are sorted by the fourth column ``Pole
Contribution'' (PC):
\begin{equation}\label{def:pole}
{\rm Pole\,\,Contribution} \equiv \frac{ \int_0^{s_0}  e^{-s/M_B^2}
\rho(s) {\rm d}s }{ \int_0^\infty  e^{-s/M_B^2} \rho(s) {\rm d}s} \,
.
\end{equation}
The pole contribution (PC) is an important quantity to check the
validity of the QCD sum rule analysis. Here, $\rho(s)$ denotes the
spectral function. It depends on the ten mixed parameters as well as
$M_B$ and $s_0$. We note that $\pi$-$\pi$ continuum which we shall
study later is not included in the pole contribution. By fixing
$s_0=1$ GeV$^2$, we show the PC values in
Table~\ref{table:parameters} for the fifteen sets. ``PC(0.5)'',
``PC(0.8)'' and ``PC(1.2)'' denote pole contribution by setting
$M_B^2 = 0.5$ GeV$^2$, $0.8$ GeV$^2$ and $1.2$ GeV$^2$,
respectively. We find that the pole contribution decreases very
rapidly as the Borel Mass increases. Since we have discussed the
convergence of OPE in the previous section, and found that the
Dim=10 and Dim=12 terms are much smaller than the Dim=6 and Dim=8
terms, and so it is only the pole contribution which gives a upper
limitation on the Borel Mass. The Borel window is wider for the
former parameter sets (1), (2), $\cdots$, and narrower for the
latter ones. It almost disappears for the set (15), whose mass
prediction is also much different from others. The Borel window
should be our working region. However, since the Borel stability is
always very good when $M_B^2 >$ 0.5 GeV$^2$, we shall keep the idea
of Borel window in mind and work in the region $0.5<M_B^2<$ 2
GeV$^2$. On the other side, we shall care more about the threshold
value $s_0$.

\begin{table}[hbt]
\caption{Values for parameters $t_i$, $\theta_i$, the mass range
$M_\sigma$, the pole contribution (PC) and the continuum amplitude
$a(t_i,\,\theta_i)$. The meaning of these quantities are given in
the text.  There are altogether fifteen sets, which are sorted by
the fourth column ``PC''. ``PC(0.5)'', ``PC(0.8)'' and ``PC(1.2)''
denote pole contribution by setting $M_B^2 = 0.5$ GeV$^2$, $0.8$
GeV$^2$ and $1.2$ GeV$^2$, respectively.}
\begin{center}
\begin{tabular}{c|ccccc|ccccc|c|ccc|c}
\hline No & $t_1$ & $t_2$ & $t_3$ & $t_4$ & $t_5$ & $\theta_1$ &
$\theta_2$ & $\theta_3$ & $\theta_4$ & $\theta_5$ & $M_\sigma$(MeV)
& PC(0.5) & PC(0.8) & PC(1.2) & a (GeV$^4$)
\\ \hline (1) & $0.03$ & $0.03$ & $0.73$ & $0.37$ & $0.24$ & $2.7$ & $3.4$ & $4.7$ & $5.5$ & $3.6$ & $510\sim580$ & 92\% & 52\% & 13\% & $ 1.2 \times 10^{-7}$
\\ \hline (2) & $0.03$ & $0.92$ & $0.75$ & $0.70$ & $0.03$ & $5.6$ & $0.80$ & $4.1$ & $2.9$ & $2.5$ & $510\sim590$ & 90\% & 46\% & 11\% & $ 5.5 \times 10^{-7}$
\\ \hline (3) & $0.25$ & $0.79$ & $0.16$ & $0.95$ & $0.22$ & $1.8$ & $1.2$ & $6.1$ & $0.44$ & $1.8$ & $510\sim600$ & 87\% & 44\% & 11\% & $ 3.6 \times 10^{-7}$
\\ \hline (4) & $0.53$ & $0.26$ & $0.93$ & $0.24$ & $0.76$ & $2.9$ & $0.40$ & $2.0$ & $2.5$ & $3.3$ & $510\sim610$ & 85\% & 41\% & 10\% & $ 1.7 \times 10^{-6}$
\\ \hline (5) & $0.74$ & $0.54$ & $0.74$ & $0.65$ & $0.67$ & $0.15$ & $3.1$ & $1.4$ & $2.7$ & $6.1$ & $520\sim640$ & 81\% & 36\% & 8\% &  $ 1.9 \times 10^{-6}$
\\ \hline (6) & $0.98$ & $0.50$ & $0.12$ & $0.33$ & $0.03$ & $2.0$ & $4.0$ & $6.3$ & $1.3$ & $1.6$ & $510\sim590$ & 82\% & 32\% & 6\% & $ 5.8 \times 10^{-8}$
\\ \hline (7) & $0.98$ & $0.42$ & $0.84$ & $0.82$ & $0.72$ & $0.095$ & $1.5$ & $3.7$ & $2.4$ & $3.0$ & $540\sim700$ & 70\% & 26\% & 6\% & $ 4.2 \times 10^{-6}$
\\ \hline (8) & $0.48$ & $0.68$ & $0.58$ & $0.96$ & $0.04$ & $1.8$ & $2.5$ & $3.0$ & $4.3$ & $3.7$ & $530\sim690$ & 70\% & 25\% & 6\% & $ 1.9 \times 10^{-6}$
\\ \hline (9) & $0.53$ & $1.0$ & $0.99$ & $0.34$ & $0.86$ & $5.6$ & $4.8$ & $5.3$ & $4.1$ & $0.076$ & $540\sim700$ & 68\% & 24\% & 5\% & $ 4.5 \times 10^{-6}$
\\ \hline (10) & $0.75$ & $0.96$ & $0.32$ & $0.12$ & $0.11$ & $4.3$ & $2.6$ & $0.93$ & $5.1$ & $2.9$ & $560\sim760$ & 57\% & 17\% & 4\% & $ 9.5 \times 10^{-7}$
\\ \hline (11) & $0.31$ & $0.81$ & $0.71$ & $0$ & $0.10$ & $4.2$ & $1.8$ & $2.8$ & $5.4$ & $5.1$ & $570\sim780$ & 55\% & 17\% & 4\% & $ 3.2 \times 10^{-6}$
\\ \hline (12) & $0.47$ & $0.40$ & $0$ & $0.46$ & $0.91$ & $0.18$ & $1.9$ & $1.9$ & $0.091$ & $0.94$ & $540\sim730$ & 58\% & 16\% & 3\% & $ 2.0 \times 10^{-7}$
\\ \hline (13) & $0.60$ & $0.26$ & $0.44$ & $0.27$ & $0.24$ & $3.3$ & $3.6$ & $0.92$ & $5.9$ & $3.7$ & $620\sim850$ & 43\% & 13\% & 3\% & $ 1.7 \times 10^{-6}$
\\ \hline (14) & $0.74$ & $0.73$ & $0.73$ & $0.32$ & $0.28$ & $1.3$ & $1.3$ & $4.6$ & $3.3$ & $5.6$ & $620\sim850$ & 42\% & 12\% & 3\% & $ 4.3 \times 10^{-6}$
\\ \hline (15) & $0.65$ & $0.55$ & $0.92$ & $0.19$ & $0.96$ & $4.9$ & $5.2$ & $4.0$ & $5.5$ & $3.3$ & $730\sim930$ & 25\% & 7\% & 2\% & $ 5.4 \times 10^{-6}$
\\ \hline
\end{tabular}\label{table:parameters}
\end{center}
\end{table}

By using these fifteen sets of numbers, we perform the QCD sum rule
analysis. There are two parameters, the Borel mass $M_B$ and the
threshold value $s_0$ in the QCD sum rule analysis. We find that the
Borel mass stability is usually good, but the threshold value
stability is not always good. We show the mass range of
$\sigma(600)$, $M_\sigma$, in Table~\ref{table:parameters}, where
the working region is taken to be $0.8$ GeV$^2<s_0<1.2$ GeV$^2$ and
$0.8$ GeV$^2< M_B^2 < 2$ GeV$^2$. We find the mass range is small
when the pole contribution (PC) is large.

\begin{figure}[hbt]
\begin{center}
\scalebox{0.6}{\includegraphics{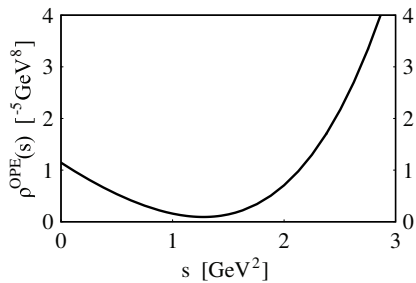}} \caption{The spectral
density $\rho(s)$ calculated by the mixed current $\eta$, as a
function of $s$. We show the results of the parameter set (02) as an
example.} \label{pic:rho02}
\end{center}
\end{figure}

\begin{figure}[hbt]
\begin{center}
\scalebox{0.6}{\includegraphics{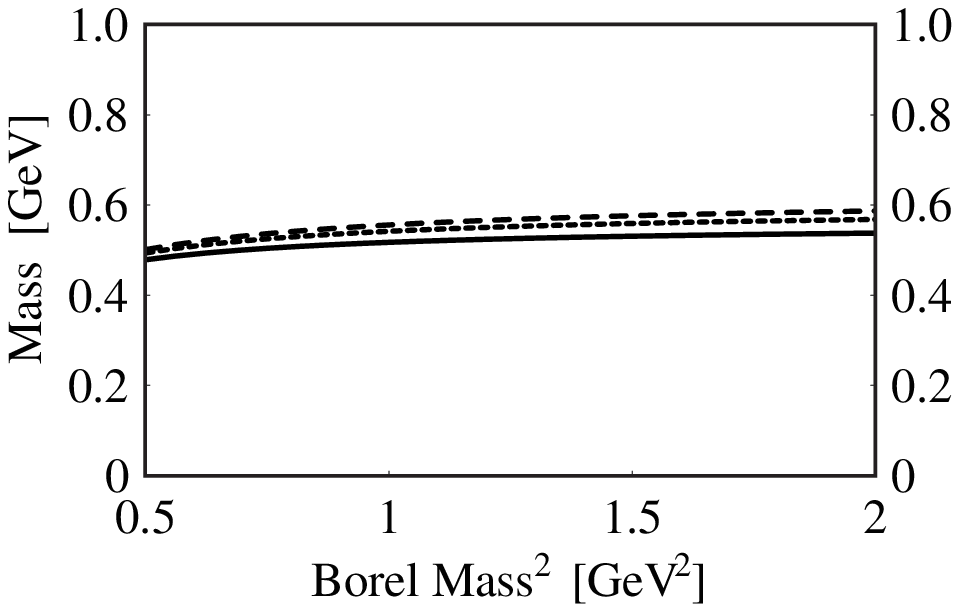}}
\scalebox{0.6}{\includegraphics{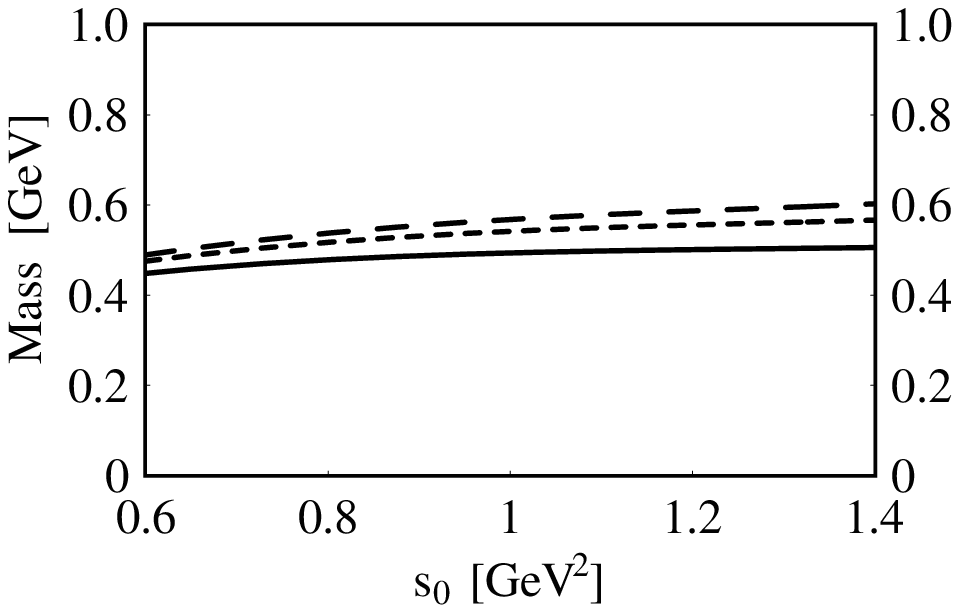}} \caption{The
extracted mass of $\sigma(600)$ as a tetraquark state calculated by
the mixed current $\eta$, as functions of the Borel mass $M_B$ and
the threshold value $s_0$. We show the results of the parameter set
(02) as an example. At the left panel, the solid, short-dashed and
long-dashed curves are obtained by setting $s_0 = 0.8,~1$ and $1.2$
GeV$^2$, respectively. At the right panel, the solid and dashed
curves are obtained by setting $M_B^2 = 0.5,~1$ and $2$ GeV$^2$,
respectively.} \label{pic:mass02}
\end{center}
\end{figure}

The parameter sets (01)-(06) lead to relatively good threshold value
stability. Taking the set (02) as an example, we show its spectral
density $\rho(s)$ in Fig~\ref{pic:rho02} as function of $s$. It is
positive definite, and has a small value around $s\sim1.2$ GeV$^2$.
Therefore, the threshold value dependence is weak around this point,
as shown in Fig.~\ref{pic:mass02} for the extracted mass as
functions of both $M_B^2$ and $s_0$. We find all the curves are very
stable in the region $0.5$ GeV$^2 < M_B^2 < 2$ GeV$^2$ and $0.6$
GeV$^2 < s_0<1.4$ GeV$^2$. From the set (02) we can extract the mass
of $\sigma(600)$ around $550$ MeV. From other good cases, we find
that the mass of $\sigma(600)$ is around $550$ MeV as well.

\begin{figure}[hbt]
\begin{center}
\scalebox{0.6}{\includegraphics{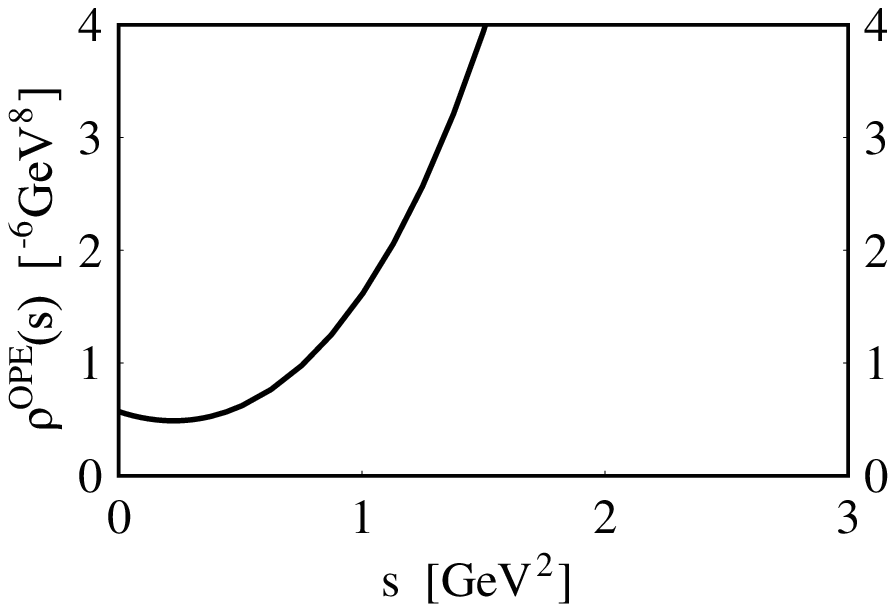}}
\scalebox{0.6}{\includegraphics{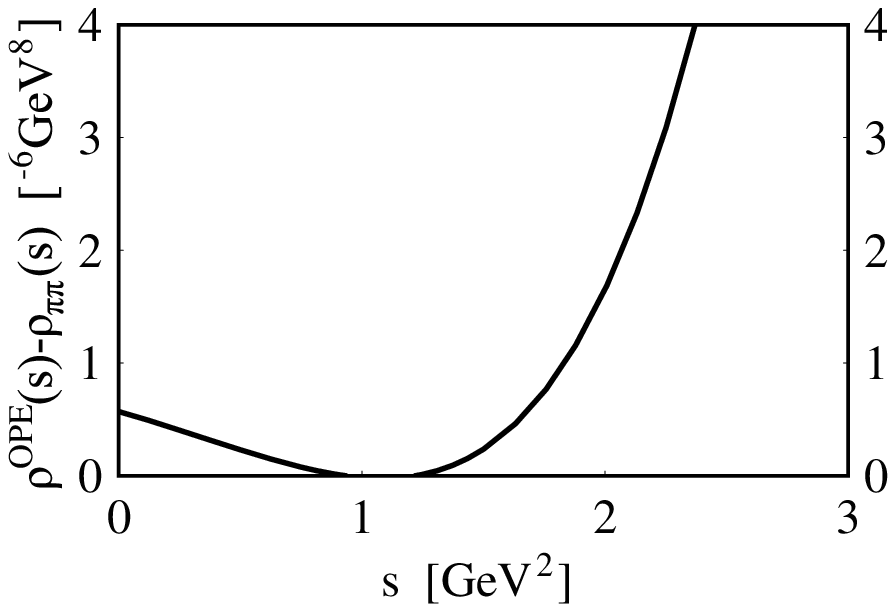}} \caption{The spectral
density $\rho(s)$ calculated by the mixed current $\eta$, as a
function of $s$. We show the results of the parameter set (13) as an
example. The left figure shows the full spectral density as given on
the left hand side of Eq.~(\ref{eq:rhonew}), while the right figure
is the one with $\rho_{\pi\pi}(s)$ subtracted.} \label{pic:rho13}
\end{center}
\end{figure}

\begin{figure}[hbt]
\begin{center}
\scalebox{0.8}{\includegraphics{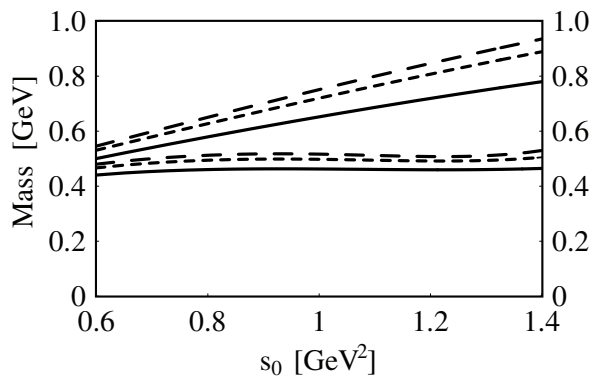}} \caption{The
extracted mass of $\sigma(600)$ as a tetraquark state calculated by
the mixed current $\eta$, as functions of the threshold value $s_0$.
We choose the parameter set (13) as an example. The solid,
short-dashed and long-dashed curves are obtained by setting $M_B^2
=0.5,~1$ and $2$ GeV$^2$, respectively. The upper three curves are
obtained without adding the contribution of the $\pi$-$\pi$
continuum in the spectral density in the phenomenological side,
while the lower three curves are obtained after adding the
contribution of the $\pi$-$\pi$ continuum.} \label{pic:mass13}
\end{center}
\end{figure}

The parameter sets (07)-(15) lead to the threshold value stability,
which is not good. Taking the set (13) as an example, we show its
spectral density in Fig.~\ref{pic:rho13} as a function of $s$ (left
figure), and the extracted mass in Fig.~\ref{pic:mass13} as a
function of $s_0$ (upper three curves). The mass increases with
$s_0$ and we cannot extract the mass from this result.
Many effects contribute to the mass dependence on the threshold
value, but for $\sigma(600)$ the $\pi$-$\pi$ continuum contribution
is probably the dominant one. Hence, we add a term
$\rho_{\pi\pi}(s)$ in the spectral function in the phenomenological
side to describe the $\pi$-$\pi$ continuum:
%
\begin{eqnarray}
\rho(s) &=& f^2_Y\delta(s-M^2_Y) + \rho_{\pi\pi}(s) + \rho_{cont} \,
. \label{eq:rhonew}
\end{eqnarray}
%
where $\rho_{cont}$ is the standard expression of the continuum
contribution except the $\pi$-$\pi$ continuum. To find an expression
for $\rho_{\pi\pi}(s)$, we introduce a coupling
\begin{eqnarray}
\lambda_{\pi\pi} &\equiv& \langle 0 | \eta | \pi^+ \pi^-
\rangle \, .
\end{eqnarray}
The correlation function of the $\pi$-$\pi$ continuum is
\begin{eqnarray}
\Pi_{\pi\pi}(p^2) &=& i \int {d^4 q \over (2\pi)^2} { i \over
(p+q)^2 - m^2_\pi + i \epsilon} { i \over q^2 - m^2_\pi + i
\epsilon} |\lambda_{\pi\pi}|^2 \, ,
\end{eqnarray}
and the spectral density of the $\pi$-$\pi$ continuum is just its
imaginary part
\begin{eqnarray}
\rho_{\pi\pi}(s) = {\rm Im} \Pi_{\pi\pi}(s) = { 1 \over 16\pi^2}
\sqrt{1-{4m_\pi^2 \over s}} |\lambda_{\pi\pi}|^2 \, .
\end{eqnarray}
We may calculate $\lambda_{\pi\pi}$ by using the method of current
algebra if we know the property of the resonance state. However,
this is not the topic of this paper. Moreover, in this paper we use
a general local tetraquark current to test the full space of local
tetraquark currents, so we again make some try and error tests, and
find that the following function leads to a reasonable QCD sum rule
result, $\lambda_{\pi\pi} \sim s$. Hence, we take the spectral
density of the $\pi$-$\pi$ continuum as
\begin{eqnarray}
\rho_{\pi\pi}(s) = a(t_i, \theta_i) s^2 \sqrt{1-{4m_\pi^2 \over s}}
\, .
\end{eqnarray}

We add the continuum contribution $\rho_{\pi\pi}(s)$ in the
phenomenological side and perform the QCD sum rule analysis. The
values of parameter $a(t_i, \theta_i)$ are listed in
Table~\ref{table:parameters}. After adding the continuum
contribution, the threshold value stability becomes much better.
Still taking the set (13) as an example, we show its spectral
density in Fig.~\ref{pic:rho13} as a function of $s$ (right figure),
and the extracted mass in Fig.~\ref{pic:mass13} as functions of
$s_0$ (lower three curves). We see that now the spectral density has
a small value around $s\sim1.1$ GeV$^2$, and the stability of the
threshold value is significantly improved.

\begin{figure}[hbt]
\begin{center}
\scalebox{0.7}{\includegraphics{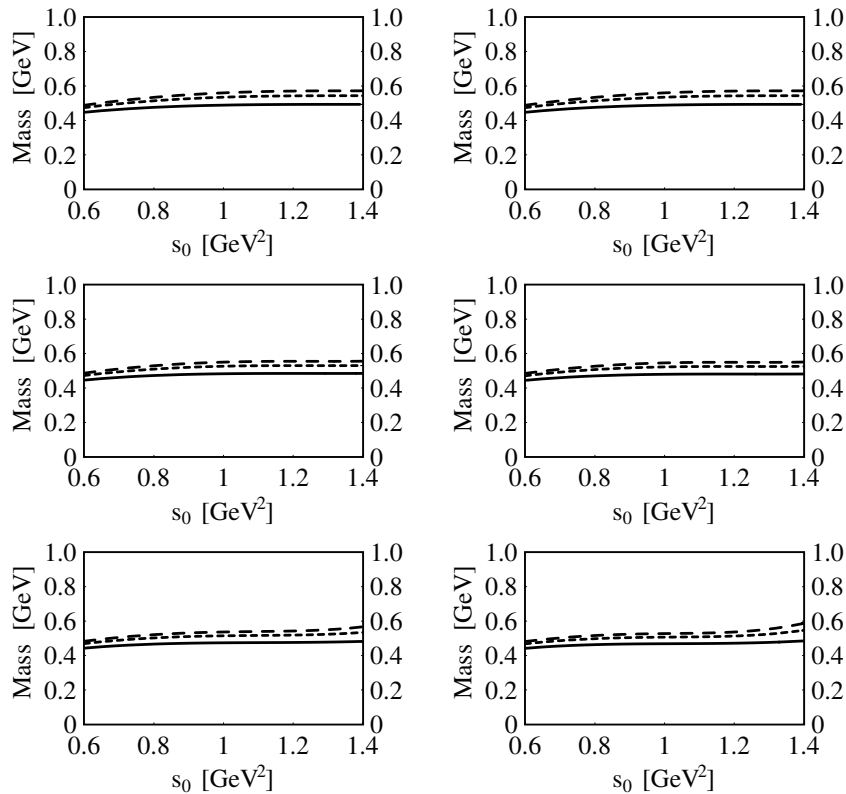}} \caption{The extracted
mass of $\sigma(600)$ as a tetraquark state calculated by the mixed
currents $\eta$, as functions of the threshold value $s_0$. We
choose the parameter sets (01), (03), (06), (09), (12) and (14). The
results are shown in sequence. The solid, short-dashed and
long-dashed curves are obtained by setting $M_B^2 = 0.5,~1$ and $2$
GeV$^2$, respectively.} \label{pic:mass}
\end{center}
\end{figure}

Hence, we made the same analysis for all the other cases. We found
all the cases are good except one, which is the case (15), where we
are not able to get the desired stability as a function of $s_0$.
The mass function has a small stability region and increases rapidly
with $s_0$. Hence, we consider this case is between the good case
and bad case, and remove it from the further analysis in this paper.
We show several results out of all the good cases in
Fig.~\ref{pic:mass}, which are obtained by using the parameter sets
(01), (03), (06), (09), (12) and (14). We list the used
$a(t_i,\theta_i)$ in Table 1 for all the cases. All the masses
behave very nicely as functions of the Borel mass and $s_0$ as shown
in Fig.~\ref{pic:mass}.  In our working region $0.8$ GeV$^2<s_0<1.2$
GeV$^2$ and $0.8$ GeV$^2<M_B^2 < 2$ GeV$^2$, all the cases lead to a
mass within the region $495$ MeV$\sim570$ MeV. From this mass range,
the mass of $\sigma(600)$ is extracted to be $530$ MeV $\pm$ 40 MeV.
%
%

\section{The Effect of Finite Decay Width}\label{sec:width}

\begin{figure}[hbt]
\begin{center}
\scalebox{0.8}{\includegraphics{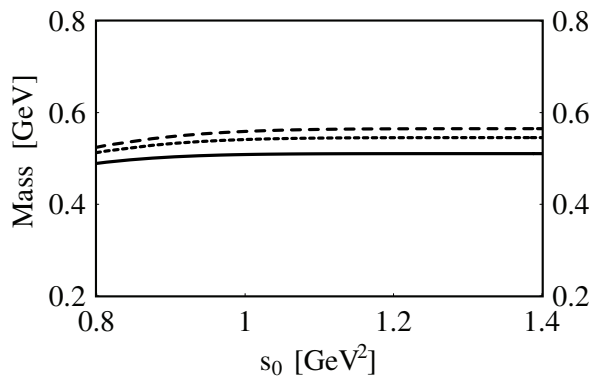}} \caption{The extracted
``mass'' considering a finite decay width. The solid, short-dashed
and long-dashed curves are obtained by setting $M_B^2 = 0.5,~1$ and
$2$ GeV$^2$, respectively.} \label{pic:massmb}
\end{center}
\end{figure}

After the $s_0$ stability has been improved, we notice now that the
mass increases systematically with the Borel mass as seen in
Fig.~\ref{pic:mass} in all the cases. We therefore try to consider a
possible reason of this systematic result. The $\sigma(600)$ meson
has a large decay width. We parametrize it by a Gaussian
distribution instead of the $\delta$-function for the $\sigma(600)$.
\begin{eqnarray}
\rho^{FDW}(s) =  {f^2_X \over
\sqrt{2 \pi} \sigma_X } \exp \big ( - { (\sqrt{s} - M_X)^2 \over 2
\sigma_X^2} \big )~.
\end{eqnarray}
The Gaussian width $\sigma_X$ is related to the Breit-Wigner decay
width $\Gamma$ by $\sigma_X = \Gamma / 2.4$. We set $\sigma_X = 200$
MeV, and $M_X = 550$ MeV, and calculate the following ``mass'':
\begin{eqnarray}
M^2(M_B, s_0) = { \int_{0}^{s_0} e^{-s/M_B^2} s \exp \big ( - {
(\sqrt{s} - M_X)^2 \over 2 \sigma_X^2} \big ) {d s \over 2 \sqrt s}
\over \int_{0}^{s_0} e^{-s/M_B^2} \exp \big ( - { (\sqrt{s} - M_X)^2
\over 2 \sigma_X^2} \big ) {d s \over 2 \sqrt s} }~.
\end{eqnarray}
We find that the obtained mass $M$ is not just 550 MeV, but
increases as $M_B^2$ increases as shown in Fig.~\ref{pic:massmb}.
Hence, the extracted mass in the QCD sum rule analysis ought to
depend on the Borel mass. The amount of the change of the extracted
mass in the QCD sum rule analysis is similar to the one found in
this model calculation. Moreover, we find that the finite decay
width does not change the final result significantly, which we have
also noticed in our previous paper~\cite{Chen:2007xr}.
%

\section{Summary}\label{sec:summary}

In summary, we have studied the light scalar meson $\sigma(600)$ in
the QCD sum rule. We have used general local tetraquark currents
which are linear combinations of five independent local ones. This
describes the full space of local tetraquark currents which can
couple to $\sigma(600)$ either strongly or weakly. We find some
cases where the stability of the Borel mass and threshold value is
both good, while in some cases the threshold value stability is not
so good. The resonance mass has an increasing trend with $s_0$,
which indicates a continuum contribution. Hence, we have introduced
a contribution from the $\pi$-$\pi$ continuum, and obtained a good
threshold value stability. The mass of $\sigma(600)$ is extracted to
be $530$ MeV $\pm$ 40MeV. Very interesting observation is that the
mass increases slightly with the Borel mass. We have made a model
calculation by taking the Gaussian width of $200$ MeV centered at
$550$ MeV and try to make a sum rule analysis. We see a similar
increase trend as seen in the QCD sum rule analysis.

The continuum contribution exists in the background of the
$\sigma(600)$ meson and it is very important to consider this fact
in the QCD sum rule analysis for exotic states. We have seen clear
tendency of the mass increase with the Borel mass after getting good
signal of the threshold dependence. The decay width of $\sigma(600)$
is related to this increase tendency. We are now trying to calculate
this by using the three-point correlation function within the QCD
sum rule approach. The present analysis is very encouraging to apply
the QCD sum rule including the continuum states for other scalar
mesons. Moreover, the continuum contribution should be important in
many other resonances such as $\Lambda(1405)$ etc, which lies in
some continuum background. In the future, we will use the QCD sum
rule analysis with continuum to study various resonances.

\section*{Acknowledgments}
This project is supported by the National Natural Science Foundation
of China under Grants No.~10625521 and No.~10721063, the Ministry of
Science and Technology of China (2009CB825200), the Ministry of
Education research Grant: Kakenhi (18540269), and the Grant for
Scientific Research ((C) No.~19540297) from the Ministry of
Education, Culture, Science and Technology, Japan.

\end{document}